# Self-Dimensioning and Planning of Small Cell Capacity in Multi-Tenant 5G Networks

P. Muñoz, O. Sallent, J. Pérez-Romero

Universitat Politècnica de Catalunya (UPC)

[pablo.munoz, sallent, jorperez]@tsc.upc.edu

*Abstract*—An important concept in the 5[th] generation of mobile networks is multi-tenancy, which allows diverse operators sharing the same wireless infrastructure. To support this feature in conjunction with the challenging performance requirements of future networks, more automated and faster planning of the required radio capacity is needed. Likewise, installing Small Cells is an effective resource to provide greater performance and capacity to both indoor and outdoor places. This paper proposes a new framework for automated cell planning in multi-tenant Small Cell networks. In particular, taking advantage of the available network data, a set of detailed planning specifications over time and space domains are generated in order to meet the contracted capacity by each tenant. Then, the network infrastructure and configuration are updated according to an algorithm that considers different actions such as adding/removing channels and adding or relocating small cells. The simulation results show the effectiveness of various methods to derive the planning specifications depending on the correlation between the tenant's and network's traffic demands.

## I. Introduction

Cellular data traffic has grown exponentially in the last few years due to the increasing popularity of new mobile devices and application services. A promising solution to satisfy this demand in the forthcoming fifth generation (5G) is based on Small Cell (SC) deployments [1]. SCs, which are more economically attractive than macrocells, provide additional capacity to the macro-cell layer, offering overlay coverage over the area of interest. The dense deployment of SCs has proven to be a cost-effective way to offer more capacity and more spectrum reuse because of their smaller cell radius [2-3].

The future 5G radio access networks (RANs) will support multi-tenancy, so that multiple mobile operators, service providers, over-the-top companies and vertical sectors can be served over the same infrastructure. In this respect, the dynamic resource provisioning between tenants has been studied in [4-5], where a central entity is responsible for allocating resources via resource slicing. This kind of solutions are intended for operating in short-term time scales. From a perspective of larger time scales, multi-tenancy poses unprecedented challenges to the owner of the shared RAN in relation to radio network planning (RNP). For example, each "tenant" has specific characteristics on its geographic and/or population coverage [6], which may also change frequently over the time. To ensure speedy and efficient deployment of services, traditional RNP has to be evolved toward new models, where SCs are considered as a key element to increase capacity.

Thanks to the Self-Organizing Network (SON) concept introduced by the 3[rd] Generation Partnership Project (3GPP) [7], traditional management tasks in cellular networks have been transformed into a set of automated functions. Self-planning is defined in [8] as the process of identifying the parameter settings of new network elements, including site locations and hardware configuration. It was included within the SON use cases defined by the Next Generation Mobile Networks (NGMN) alliance [9]. To meet the envisioned capacity of 5G networks, the concept of "Self-" has to be applied to the dimensioning, planning and deployment of SCs. By making these activities more dynamic, faster and automatic, capital and operational expenditures can be reduced and network performance improved. Thus, the new RNP functions will operate at shorter time scales than today, involving a set of decision-making processes that can be triggered by various events. The diversity of these events will also be much greater in the future 5G networks, ranging from call traces and cell counters crossing a given threshold to the arrival of new tenants. In addition, the decision-making processes will manage a wider range of cost-efficient solutions, taking advantage of the flexibility of SCs and being integrated with the optimization tasks to avoid suboptimal network configuration and inconsistencies.

Taking into account the gaps in the open literature, in this paper, a novel framework that applies the SON principles to the RNP problem is proposed to meet the challenging requirements of 5G. In particular, this work tackles the RNP problem in the context of 5G considering the following: (i) capacity provisioning will be one of the most challenging issues for network equipment providers; (ii) the multi-tenancy feature will introduce additional complexity to the planning process since multiple capacity conformance specifications over the spatial and temporal domains will be defined; and (iii) SCs will be considered as the main driver to satisfy the traffic demand because they are a less expensive and easy-to-deploy alternative to macrocells. Deploying SCs entails strong implications in the way that the spectrum planning (or channel allocation) is carried out, since the newly deployed SCs will interfere other co-channel SCs. Nevertheless, the spectrum planning problem can be solved in a more localized fashion than in macrocell deployments, because of the smaller size of SCs and the usage of high carrier frequencies, which facilitate

This work has been accepted for publication in the IEEE Transactions on Vehicular Technology. Citation information: DOI 10.1109/TVT.2018.2793418. Copyright (c) 2018 IEEE. Personal use of this material is permitted. However, permission to use this material for any other purposes must be obtained from the IEEE by sending a request to pubs-permissions@ieee.org. This work has been supported by the EU funded H2020 5G-PPP project 5G ESSENCE under the grant agreement 761592, by the Spanish Research Council and FEDER funds under RAMSES grant (ref. TEC2013-41698-R) and by the Spanish Ministry of Economy and Competitiveness (ref. FJCI-2014-19815).

extensive spatial reuse. Thus, the proposed framework advocates a more systematic and efficient way of RNP without requiring vast computational resources. In this respect, when network measurements, call traces and/or geo-located data are available, the models to characterize network performance and traffic estimation can exploit them. The framework also reduces the gap between planning and optimization work by defining a unified approach where the potential solutions are intended to satisfy the traffic demand while minimizing capacity overprovisioning in the network.

The remainder of this paper is organized as follows. In Section II the literature related to RNP is discussed. In Section III the system model is presented. Section IV describes the proposed architecture and analyzes the multi-tenancy feature from a planning perspective. Results are discussed in Section V. Finally, Section VI provides some concluding remarks.

## II. RELATED WORK

The RNP problem has been widely investigated in macrocell scenarios [10-11]. With the arrival of SCs, the RNP problem has been addressed in the context of Heterogeneous Networks, where the cross-tier interference between macrocells and SCs has been crucial for an efficient deployment [12-17]. However, focusing on the SC-layer, there is a lack of studies on the joint optimization of SC location planning and spectrum planning. In general, spectrum planning (or channel allocation) is seen as a separate issue that is solved only when the location of the SCs is already known.

RNP as an optimization problem has commonly been approached as a two-phase process, where the first phase (also referred to as dimensioning) has to determine the minimum number of cells to satisfy signal coverage, system capacity and cost constraints, while the second phase tries to find the optimal cell locations [10][12][18-19]. Eliminating redundant base stations can be considered in the method as a third phase [11], or it can be the fundamental basis of the planning method [16]. Another approach is based on iteratively adding a cell which has the highest increment of an utility function among the set of candidate locations until network capacity reaches a target value [20-21]. Note that, in this case, the a priori dimensioning phase is not required. In general, the number of cells to deploy is estimated as the total traffic demand of the considered area divided by the average capacity of a cell. In SC networks, the SCs can use different number and different set of channels, which also may change over time. Consequently, the estimation of cell capacity is not as simple as in macrocell networks, where the frequency reuse is typically one. This issue obviously affects the accuracy of the estimation of the number of cells. Thus, to avoid relying on this number, in this paper, the iterative approach of adding SCs has been selected.

The success of RNP depends to some extent on the model that is used to characterize the network performance, typically Monte-Carlo simulations [10][22] or stochastic processes [23-25]. These models can be utilized to develop either proactive or reactive planning strategies. Recently, the energy consumption of base stations has also been modeled to provide energy-efficient planning solutions [20][23][26-27]. From a more practical perspective, existing commercial planning tools (such as those mentioned in [19]) provide operators with multitude of capabilities for efficient RNP, including realistic traffic maps and accurate propagation models. However, there is very little open information about methodology and principles due to the confidential and proprietary nature of these solutions. Lastly, the RNP of multi-tenant networks has received little attention in the literature [28].

From the perspective of 5G, network slicing represents a fundamental feature to accommodate traffic demands in multi-tenant networks without significantly increasing operational and infrastructure costs [29]. Specifically, network slicing consists of partitioning a common physical infrastructure into multiple logical networks. The challenges that arise from introducing this feature in 5G networks have been addressed in several works [29-30]. The architectural issues for enforcing slices in mobile networks have been studied in [31-32], while the algorithmic aspects have been investigated in [33]. In that work, an algorithm for dynamic resource sharing across slices taking into account user association decisions has been proposed. To enable network slicing in mobile networks, various technologies such as Network Function Virtualization (NFV), Software-Defined Networks (SDN) and cloud computing have been considered [34]. In this respect, the allocation of virtualized network functions, the network programmability and the centralized coordination are key aspects for the adoption of network slicing and the diverse requirements of 5G mobile networks.

With network slicing, a much more efficient utilization of network resources is achieved in multi-tenant networks. However, when traffic volumes increase and the current infrastructure is not enough to meet the required capacity, investing on new resources remains as the only solution to this issue. While the above references focus on network slicing to optimize the resource usage during the network operation, this paper proposes a framework to update the network infrastructure during the (re-)planning phase. Such a framework was initially introduced in [35], where the SC location problem and spectrum planning were approached as a joint problem. This paper further develops this initial work introducing the following contributions and novelties:

- A novel implementation of the SC planning and spectrum allocation processes is considered. Specifically, when a new SC has to be deployed, the channel allocation process is performed through an iterative process to try a variable number of allocated channels.
- New scenarios are adopted to evaluate a greater variety of planning actions, such as removing a channel in a SC.
- An exhaustive temporal analysis of the proposed planning strategies is performed to evaluate the re-planning phase. This phase takes place after deploying a new tenant and it is especially effective when the tenant's demand is a priori unknown.
- The proposed capacity planning method is evaluated against the state-of-the-art, where the spectrum allocation is commonly seen as an independent function that is executed only when the location of the SCs is already known.

## III. SYSTEM MODEL

Let us assume a scenario where a certain infrastructure provider owns a RAN comprised of SCs. The SCs are intended to meet the high capacity requirements in localized areas. The provider offers at time $t$ such a RAN to a certain number $M^{(t)}$ of tenants, so that the tenants' customers can get access to the tenant's service. Let denote as $D_{i,m}^{(t)}$ the traffic demand (in Mbps) of tenant $m$ in SC $i$ at time $t$, calculated as the sum of the traffic from all the users attached to SC $i$. The tenants' traffic $D_{i,m}^{(t)}$ can be aggregated into a new variable, $D_i^{(t)}$, which provides the total traffic demand in SC $i$, i.e.:

$$D_i^{(t)} = \sum_{m=1}^{M^{(t)}} D_{i,m}^{(t)}. \qquad (1)$$

Let also $d_{u,m}^{(t)}$ be the traffic demand (in Mbps) of tenant $m$ in the $u^{\text{th}}$ pixel of the scenario ($u \in U$). The metric $d_u^{(t)}$ is computed in a similar way to (1) in order to determine the total traffic at the pixel-level.

The geographical area of interest is divided into a set $U$ of grid points, called pixels. A subset $U_C \subseteq U$ of these locations are candidate site locations for SCs. A typical placement of a SC site is below rooftops. However, in many cases, the selection of potential sites depends on how easily they can be acquired and backhauled, e.g. if there exists line of sight to a nearby hub. In this respect, an adequate filter of unaffordable site locations to determine the subset $U_C$ will reduce computational complexity of the later planning process. Finally, let $U_S^{(t)} \subseteq U_C$ be the subset of site locations with deployed SCs at time $t$.

The transmit power and the allocated bandwidth of the $i^{\text{th}}$ SC are denoted by $P_i^{(t)}$ and $B_i^{(t)}$, respectively. The transmit power is configured such that it provides a certain Signal-to-Interference-plus-Noise Ratio (SINR) at the targeted coverage range [36]. With respect to the carrier frequency, SCs are assumed to be deployed in higher frequencies than the 1~2 GHz, such as e.g. the 5 GHz considered by the 3GPP as a feasible solution [36]. The frequency band is partitioned into a set $F = \{f_1, ..., f_K\}$ of $K$ orthogonal channels of bandwidth $B$. The subset of channels allocated to SC $i$ at time $t$ is given by $F_i^{(t)} \subseteq F$. Therefore, the total bandwidth allocated to SC $i$ is expressed as $B_i^{(t)} = |F_i^{(t)}| \cdot B$, where $|\cdot|$ denotes cardinality.

The capacity of SC $i$ is given by:

$$C_i^{(t)} = B_i^{(t)} \cdot \overline{SE}_i^{(t)}, \qquad (2)$$

where $\overline{SE}_i^{(t)}$ represents the average spectral efficiency achievable at SC $i$. In general, the spectral efficiency depends on the radio access technology and the SINR conditions.

To determine areas in the network with a lack of capacity and areas with spare capacity, the required bandwidth becomes a key metric in the planning process. Specifically, this metric at the pixel-level can be determined from the traffic demand and the spectral efficiency as follows:

$$B_{i,u}^{\prime(t)} = \frac{d_u^{(t)}}{SE_i^{(t)}(u)}. \qquad (3)$$

In a similar way, $B_i^{\prime(t)}$ represents the required bandwidth on a cell basis.

IV. FUNCTIONAL ARCHITECTURE

This section focuses on elaborating a reference framework for multi-tenant management from the perspective of network planning. The proposed model is depicted in Fig. 1. The network, represented in the bottom of the figure, is characterized by the network configuration, which is given by $U_S^{(t)}$ and $F_i^{(t)}$. The network performance can be seen as a source of relevant information for the planning process. In particular, it provides a collection of metrics related to the past and actual traffic demand and also to the quality of the offered services. The information can be given at either the SC-level or pixel-level. In the former case, the metrics are derived from cell counters and they are typically known as Key Performance Indicators (KPIs). In the latter case, the information is derived from call traces, which contain geo-located measurements from users.

The functional architecture of Fig. 1 includes two main entities described in the following. These entities can be part of the management systems such as the Element Manager (EM) or the Network Manager (NM) [37].

A. *Multi-tenancy management entity*

The multi-tenancy management entity acts as an interface between the tenants and the network planning tool of the network provider. From the perspective of planning, the SLA defines the contracted capacity $\hat{A}_m$ (in Mbps) that tenant $m$ demands to the network provider. Normally, it is expressed in terms of aggregate (or average) values over relatively coarse time and space scales. The SLA may also include some other

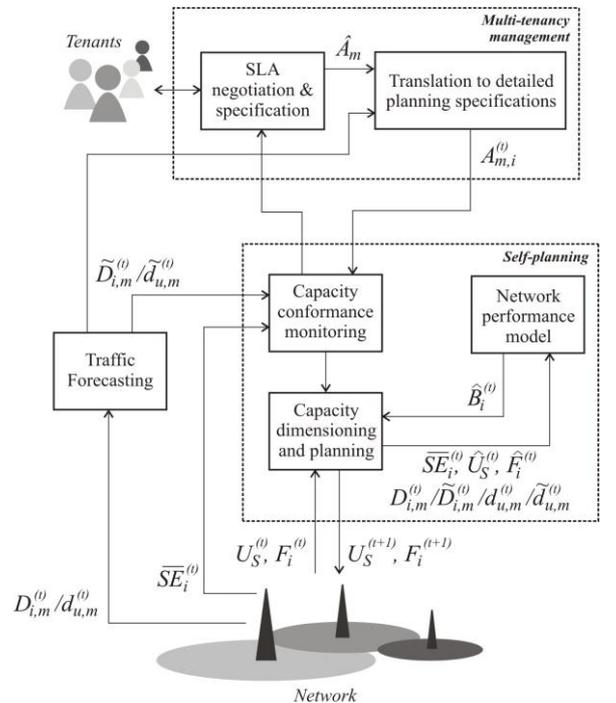

*Fig. 1. Functional architecture.*

guarantees, e.g. related to QoS metrics.

For network planning purposes, the SLA has to be expressed in smaller time and space scales that can be more easily used when taking planning decisions. In particular, the contracted capacity $\hat{A}_m$ is translated into a set of detailed planning specifications $A_{m,i}^{(t)}$ that depend on SC $i$ and time $t$. To do this, the current or predicted traffic demand in the network can be employed. This process, which ensures that the contracted capacity is provided, depends exclusively on the network provider's side. As a consequence, the SLA is simplified and tenants are excluded from gaining a detailed picture of the network infrastructure.

Each time a set of detailed planning specifications are generated, the multi-tenancy management entity sends this information to the self-planning entity, which will use them to determine whether the infrastructure needs to be updated or not. This situation typically occurs when a new tenant is aggregated or an ongoing tenant updates the SLA. In the former case, the traffic demand of the tenant is unknown, while in the latter case it depends on the geographical area where the contracted capacity is modified. For example, if an ongoing tenant extends its service coverage (e.g. according to business plans) to a new geographical area that is owned by the infrastructure provider, the traffic demand of the tenant in this area is unknown. On the contrary, if the tenant modifies the contracted capacity (i.e. due to an increase in traffic demand) within the limits of the current service coverage area, the temporal and spatial distributions of the traffic demand are already known in this case.

Depending on whether the traffic demand of the tenant is unknown or not, the detailed planning specifications are generated in a different way.

In the first case (i.e. traffic demand is unknown), $A_{m,i}^{(t)}$ is calculated based on the temporal and spatial variations of the traffic demand from other tenants. Specifically, the temporal variation of the traffic demand is mainly given by the traffic fluctuations that take place over one day's time. Such a temporal pattern is typically repeated over different days. An example sequence of total tenants' traffic demand $D^{(t)}$ during several days is illustrated in Fig. 2, where $T$ stands for the one-day period and $t_B$ is the busy hour. Based on this, let $A_m^{(t_B)}$ be the detailed planning specification at the busy hour, which can be estimated from $\hat{A}_m$ and from the time variations of the other tenants' traffic demand as follows:

$$A_m^{(t_B)} = \hat{A}_m \cdot \frac{D^{(t_B)}}{\bar{D}}. \qquad (4)$$

Regarding the spatial variations of the traffic demand, the contracted capacity at the busy hour $A_m^{(t_B)}$ is distributed among the number $|U_S^{(t)}|$ of deployed SCs taking into account the following condition:

$$A_m^{(t_B)} = \sum_{i \in U_S^{(t)}} A_{m,i}^{(t_B)}, \qquad (5)$$

where $A_{m,i}^{(t_B)}$ is the contribution of the contracted capacity in SC $i$. Depending on the spatial correlation that can be expected between the tenant's traffic demand and the actual network's traffic demand, the detailed planning specifications per cell for tenant $m$ can be formulated in different ways:

- *Uniform distribution*. In case that the spatial traffic demand of the new tenant is unknown, an even distribution of traffic is assumed. Estimation can be conducted at the SC-level (6) or pixel-level (7):

$$A_{m,i}^{(t_B)} = \frac{A_m^{(t_B)}}{|U_S^{(t)}|}, \qquad (6)$$

$$A_{m,u}^{(t_B)} = \frac{A_m^{(t_B)}}{|U|}, \qquad (7)$$

where $A_{m,u}^{(t_B)}$ stands for the contracted capacity at the $u^{th}$ pixel and $|U|$ is the total number of pixels in the area. If uniform distribution at the SC-level is adopted, the values of $A_{m,u}^{(t_B)}$ at pixel $u$ are obtained considering a uniform distribution of traffic within the service area of the corresponding SC. On the other hand, when estimation is at the pixel-level, the value of $A_{m,i}^{(t_B)}$ in SC $i$ is obtained from aggregating the values of $A_{m,u}^{(t_B)}$ only for pixels served by SC $i$. Note that this value may be different to that obtained from using (6).

- *Correlated distribution*. In case that the correlation between the traffic demand for the new tenant and the already existing tenants is expected, areas with higher traffic demand of other tenants will receive a greater contribution of $A_m^{(t_B)}$. Such an estimation can also be conducted at either SC- or pixel-level. In the former case, using the information on KPIs that measure $D_i^{(t_B)}$ in SC $i$ as an estimation of the spatial traffic demand of tenant $m$, the detailed planning specification is given by:

$$A_{m,i}^{(t_B)} = A_m^{(t_B)} \cdot \frac{D_i^{(t_B)}}{\sum_{p \in U_S^{(t)}} D_p^{(t_B)}}. \qquad (8)$$

In the latter case, the traffic measurements at the pixel-level are taken from call traces that provide geo-located information for each user in an automatic way. Thus, the specification is calculated as:

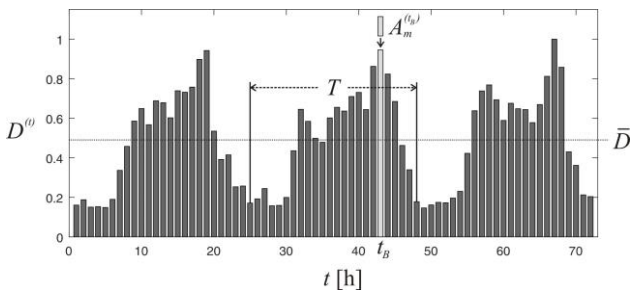

*Fig. 2. Example of normalized traffic demand over three days.*

$$A_{m,u}^{(t_B)} = A_m^{(t_B)} \cdot \frac{d_u^{(t_B)}}{\sum_{v \in U^{(t)}} d_v^{(t_B)}}, \qquad (9)$$

where $d_u^{(t_B)}$ is the total traffic demand in the $u^{\text{th}}$ pixel of the scenario.

In the second case (i.e. traffic demand is known), the temporal and spatial variations of the traffic demand of tenant $m$ are considered for the generation of the detailed planning specifications over such domains. In particular, if the traffic demand is given at the SC-level, $D_i^{(t_B)}$ is replaced by $D_{i,m}^{(t_B)}$ in (8). If, on the contrary, the traffic demand is given at the pixel-level, $d_u^{(t_B)}$ is replaced by $d_{u,m}^{(t_B)}$ in (9).

### B. Self-planning entity

According to Fig. 1, the detailed planning specifications, $A_{m,i}^{(t)}$, are used by the self-planning entity, whose aim is twofold. The first objective is to check whether or not the deployed network capacity fits the tenants' demand. The second is to provide the required changes in the network layout and channel allocation, given by $U_S^{(t+1)}$ and $F_i^{(t+1)}$ respectively, in case there is a lack of capacity. The self-planning entity follows an automated approach characterized by running an iterative process that is executed during the network operation assuming that a set of SCs have already been deployed. Thus, the currently deployed infrastructure is incrementally adapted to the evolving tenants' requirements to make capacity expansion smoother, less costly and faster. The proposed approach is applied to smaller regions that are covered by a subset (or cluster) of SCs, so that dimensioning and planning tasks are accelerated and simplified. As a result, dimensioning and planning can be regarded as an automated function that can be easily integrated into the SON framework. This approach contrasts with the traditional planning, which employs longer timescales to accommodate larger capacity needs over the whole network's geographical area. Nevertheless, both traditional and SON-based planning are complementary approaches to drive network expansion at different time scales.

#### 1) Capacity conformance monitoring

This module watches over the network to determine when the network infrastructure has to be reconfigured in order to meet the tenant's traffic demand while minimizing over-provisioning. The required bandwidth in the SCs can vary due to high tenant's actual traffic demand $D_{i,m}^{(t)}$, the addition or removal of new tenants (i.e. variations in $M^{(t)}$) or changes in the planning specification $A_{m,i}^{(t)}$. Also, if the process is executed proactively, the predicted traffic demand can be considered. To this end, the *traffic forecasting* entity provides the predicted traffic demand in time $t$ at the SC-level (which can be computed from historical data using statistical models) as input in the self-planning entity. The proactive response is key as long as the deployment of new infrastructure may require substantial time compared to the evolution of the traffic demand. Thus, the traffic forecasting entity predicts the traffic growths on a relatively long-term time scale (e.g. weeks, months). As a result, the system is able to anticipate the need for more SCs and/or spectrum.

The capacity conformance is conducted in terms of the required bandwidth $\hat{B}_i^{(t)}$ by SC $i$, which can be estimated as:

$$\hat{B}_i^{(t)} = \frac{1}{\overline{SE}_i^{(t)}} \sum_{m=1}^{M^{(t)}} \min\left(D_{i,m}^{(t)}, A_{m,i}^{(t)}\right). \qquad (10)$$

In case of working with variables at the pixel-level, the translation to the SC-level is a simple aggregation of data per cell. If the traffic demand of tenant $m$ at SC $i$ is below the SLA's planning specification, $D_{i,m}^{(t)}$ is used to provide cost-effective dimensioning, since the SC's bandwidth would fit the actual required bandwidth. If, on the contrary, the traffic demand exceeds the SLA's planning specification, the required bandwidth is then limited by $A_{m,i}^{(t)}$. To compute $\hat{B}_i^{(t)}$ in practice, variable $\overline{SE}_i^{(t)}$ can be estimated from the transmitted data volume $v_i^{(t)}$ (measured in *bits*) under full-buffer conditions and the amount of resource elements $n_i^{(t)}$ (measured in $s \cdot Hz$) that have been used for transmission in SC $i$, i.e.:

$$\overline{SE}_i^{(t)} = \frac{v_i^{(t)}}{n_i^{(t)}}. \qquad (11)$$

With respect to variable $D_{i,m}^{(t)}$, it can be estimated as follows:

$$D_{i,m}^{(t)} = \frac{v_{i,m}^{(t)}}{T}, \qquad (12)$$

where $v_{i,m}^{(t)}$ is the total data volume (in *bits*) transferred in SC $i$ during time $T$ (see Fig. 2) for tenant $m$.

The bandwidth of SC $i$, $B_i^{(t)}$, is dimensioned so that the required bandwidth $\hat{B}_i^{(t)}$ is satisfied at the busy hour $t_B$, which is calculated as:

$$t_B = \arg\max_\tau \left(\hat{B}_i^{(\tau)}\right), \quad \tau = t - T + 1, ..., t. \qquad (13)$$

According to this, the *capacity conformance monitoring* module triggers the *capacity dimensioning and planning* module if the following condition is fulfilled for any of the deployed SCs in $L$ consecutive periods of $T$ duration:

$$\hat{B}_i^{(t_B)} > \alpha \cdot \left|F_i^{(t_B)}\right| \cdot B \qquad (14)$$

where $\alpha \in [0,1]$ is an adjustable parameter that determines the ability to support some variations in the traffic demand with respect to the estimated value. Note that an increase in $\hat{B}_i^{(t_B)}$ does not always trigger the condition in (14), since there can be spare capacity in the SCs. This situation can occur when: (i) the network capacity has been intentionally overprovisioned; (ii) the traffic demand has been overestimated when planning tenants for the first time; (iii) the traffic demand of any tenant has decreased over time; or (iv) a certain SLA has been modified.

Lastly, note that, if the total traffic demand of any tenant exceeds the contracted capacity, the *capacity conformance*

*monitoring* module should communicate the multi-tenancy management entity the need of reviewing (or negotiating) the SLA in order to meet the traffic demand.

*2) Capacity dimensioning and planning*

This module aims to determine the optimal solution (i.e. an updated RAN) to cope with the varying traffic demand. A candidate solution is represented by $\hat{U}_S^{(t)}$ and $\hat{F}_i^{(t)}$, which represent a modified version of the actual network deployment and spectrum allocation, respectively. The required bandwidth of the candidate solution, $\hat{B}_i^{(t)}$, is obtained from the *network performance model*, which emulates the behavior of the network with a certain layout and configuration. Unlike the SC bandwidth (measured in steps of $B$ MHz), $\hat{B}_i^{(t)}$ is a continuous variable that depends on the traffic demand and the spectral efficiency.

The dimensioning and planning is modeled as an iterative process, initiated after satisfying (14), where a set of conditions are sequentially checked at each time step in order to trigger specific planning actions (i.e. adding/removing a channel and deploying/relocating a SC). Such actions are accumulated during the planning process and, after that, the infrastructure provider is notified about the changes in the network to be implemented. The execution of planning actions depends on the limited budget of possible network changes that can be taken in a specific period. This means that the infrastructure provider is responsible for deciding when communicating the changes to the infrastructure deployment team according to other regulation and economic factors. Such a problem has been studied in [28].

The planning process is summarized in Algorithm 1, where $N_{max}^{SC}$ is the maximum number of SCs that can be deployed in the area of interest, $K_{max}$ is the maximum number of channels that can be allocated in a SC and $\beta, \gamma \in [0,1]$ are adjustable parameters. In detail, in steps 1-20, the planning process focuses on extending the capacity in areas with a lack of capacity, while in steps 21 to 31, this process aims at minimizing the capacity overprovisioning. Note that actions such as removing channels or SCs where traffic has decreased significantly may result in reduced interference, increased quality and/or capacity. During the execution of this process, a certain planning action (e.g. adding a channel) can be canceled due to the execution of the opposite action (e.g. removing a channel) depending on the actions carried out between the two (e.g. a channel added in steps 1-6 may no longer be needed if a SC is later on added in steps 7-20). The channel selection in a SC is performed so that the SC-to-SC distance between the given SC and the closest neighboring SC using the same channel is the maximum possible. This process is summarized in Algorithm 2. In addition, when a planning action is selected the *network performance model* is launched to obtain the value of required bandwidth, $\hat{B}_i^{(t)}$, corresponding to the new network configuration. In case a new SC has to be deployed, $\hat{B}_i'^{(t)}$ is calculated for each candidate site in the area of interest. Then, the site with the lowest required bandwidth is selected.

| Algorithm 1 Capacity dimensioning and planning | |
|---|---|
| 1: | // Adding a channel |
| 2: | **While** $\exists j \mid \hat{B}_j^{(t_B)} > \alpha \cdot \left|\hat{F}_j^{(t)}\right| \cdot B$ **and** $\left|\hat{F}_j^{(t)}\right| < K_{max}$ |
| 3: | Set $\left|\hat{F}_j^{(t)}\right| = \left|\hat{F}_j^{(t)}\right| + 1$; |
| 4: | $\hat{F}_j^{(t)} = $ Channel_Selection($\hat{U}_S^{(t)}, \{\hat{F}_i^{(t)}\}, j$); |
| 5: | $\hat{B}_i^{(t_B)} = $ Network_Performance_Model($\hat{U}_S^{(t)}, \{\hat{F}_i^{(t)}\}$) $\forall i \in \hat{U}_S^{(t)}$; |
| 6: | **End** |
| 7: | // Deploying a SC |
| 8: | **While** $\exists j \mid \hat{B}_j^{(t_B)} > B \cdot \dfrac{\left|U_S^{(t)}\right|}{N_{max}^{SC}/K_{max}}$ **and** $\left|U_S^{(t)}\right| < N_{max}^{SC}$ |
| 9: | $k = 0$; // $k$: number of allocated channels |
| 10: | **Do** |
| 11: | Set $k = k+1$; |
| 12: | **For all** $x \in \hat{U}_C^{(t)}$ **do**: |
| 13: | Set $\hat{U}_S'^{(t)} = \hat{U}_S^{(t)} \cup \{x\}$; |
| 14: | $\hat{F}_x'^{(t)} = $ Channel_Selection($\hat{U}_S'^{(t)}, \{\hat{F}_i^{(t)}\}, x$); |
| 15: | $\hat{B}_i'^{(t_B)} = $ Network_Performance_Model($\hat{U}_S'^{(t)}, \ldots \{\{\hat{F}_i^{(t)}\}, \hat{F}_x'^{(t)}\}$) $\forall i \in \hat{U}_S'^{(t)}$; |
| 16: | **End For** |
| 17: | Select $x^*$ with objective: $\min \sum_{i \in \hat{U}_S'^{(t)}} \hat{B}_i'^{(t_B)}$; |
| 18: | **While** $\hat{B}_{x^*}'^{(t_B)} > \alpha \cdot k \cdot B$ **and** $k < K_{max}$; |
| 19: | Set $\hat{U}_S^{(t)} = \hat{U}_S^{(t)} \cup \{x^*\}$, $\hat{U}_C^{(t)} = \hat{U}_C^{(t)} \setminus \{x^*\}$, $\hat{F}_{x^*}^{(t)} = \hat{F}_{x^*}'^{(t)}$; $\hat{B}_i^{(t_B)} = \hat{B}_i'^{(t_B)}$ $\forall i \in \hat{U}_S^{(t)}$ |
| 20: | **End** |
| 21: | // Removing a channel |
| 22: | **While** $\exists j \mid \hat{B}_j^{(t_B)} < \beta \cdot \left(\left|\hat{F}_j^{(t)}\right|-1\right) \cdot B$ **and** $\left|\hat{F}_j^{(t)}\right| > 1$ |
| 23: | Set $\left|\hat{F}_j^{(t)}\right| = \left|\hat{F}_j^{(t)}\right| - 1$; |
| 24: | $\hat{F}_j^{(t)} = $ Channel_Selection($\hat{U}_S^{(t)}, \{\hat{F}_i^{(t)}\}, j$); |
| 25: | $\hat{B}_i^{(t_B)} = $ Network_Performance_Model($\hat{U}_S^{(t)}, \{\hat{F}_i^{(t)}\}$) $\forall i \in \hat{U}_S^{(t)}$; |
| 26: | **End** |
| 27: | // Removing a SC |
| 28: | **While** $\exists j \mid \hat{B}_j^{(t_B)} < \gamma \cdot B$ |
| 29: | Set $\hat{U}_S^{(t)} = \hat{U}_S^{(t)} \setminus \{j\}$, $\hat{U}_C^{(t)} = \hat{U}_C^{(t)} \cup \{j\}$; |
| 30: | $\hat{B}_i^{(t_B)} = $ Network_Performance_Model($\hat{U}_S^{(t)}, \{\hat{F}_i^{(t)}\}$) $\forall i \in \hat{U}_S^{(t)}$; |
| 31: | **End** |

| 32: | Set $U_S^{(t)} = \hat{U}_S^{(t)}$, $U_C^{(t)} = \hat{U}_C^{(t)}$ ; $F_i^{(t)} = \hat{F}_i^{(t)}$, $\forall i \in \hat{U}_S^{(t)}$ |

Algorithm 1 is designed such that the deployment of new SCs is carried out gradually as the traffic demand grows and new channels are allocated. According to this, as the number of deployed SCs gets closer to the saturation point (given by $\left|U_S^{(t)}\right| = N_{max}^{SC}$), the threshold in the first condition of step 8 that is used to deploy new SCs approaches the value of maximum amount of allocable bandwidth in a SC (i.e. $K_{max} \cdot B$).

Regarding the computational complexity of Algorithm 1, it is worth highlighting that, as long as the dimensioning and planning are rather long-term processes, computational complexity is not a first-order requirement to consider. However, to limit the complexity when the number of SCs increases, the considered geographical area could be divided into smaller regions so that Algorithm 1 is applied to each of them independently.

From an economic perspective, each kind of planning action entails a different cost for the infrastructure provider. In particular, adding or removing channels represents the cheapest solution since it can be executed remotely. On the opposite side is the deployment of new SCs, which requires investing on new infrastructure. The infrastructure provider can be interested in balancing the priority of these two planning actions according to its financial objectives. This can be done in the proposed algorithm by tuning $N_{max}^{SC}$, which is the parameter that has a direct impact on the deployment costs.

| **Algorithm 2** Channel selection |
|---|
| 1: **Inputs**: $\hat{U}_S^{(t)}$, $\left\{\hat{F}_i^{(t)}\right\}$, $x$: targeted SC; |
| 2: **Initialize**: $\hat{F}_x^{(t)} = \varnothing$; |
| 3: **Do**: |
| 4:     Calculate $s(x, y)$; // $s$: distance between $x$ and $y$ |
| 5:     $s_j(x, y) = \begin{cases} s(x, y) & \text{if } f_j \subseteq \left(\hat{F}_x^{(t)} \cap \hat{F}_y^{(t)}\right) \\ +\text{Inf} & \text{otherwise} \end{cases}$ |
| 6:     $s_j(x) = \min_y s_j(x, y)$, $\forall f_j \in F$; |
| 7:     $i = \arg\max_j s_j(x)$; |
| 8:     Set $\hat{F}_x^{(t)} = \hat{F}_x^{(t)} \cup \{f_i\}$; $F = F \setminus \{f_i\}$; |
| 9: **While** $\left|\hat{F}_x^{(t)}\right| < k$; |

The parameter $K_{max}$ is also adjustable and it determines the interference levels that are allowed in the network. For example, by setting the maximum value, all the SCs can use all the channels. However, a high value of this parameter may not be recommended as it would result in excessive interference levels, making the solution spectrally inefficient. Likewise, a too low value of this parameter should be avoided since it entails a waste of spectrum and an increased cost due to a faster deployment of SCs.

The parameters $\alpha$ and $\beta$ determine the amount of spare capacity that is retained by the infrastructure provider in the SCs e.g. to absorb eventual peaks of traffic demand. They are jointly configured to avoid recursive channel allocations and releases in the SCs.

Lastly, the parameter $\gamma$ establishes the sensitivity of the planning action related to relocation of SCs that support marginal amount of traffic. This situation happens, for example, when a tenant's contract expires leaving a large amount of spare capacity in the SCs.

*3) Network performance model*

To evaluate the candidate solutions, a *network performance model* is required. The objective of this model is to compute the required bandwidth in the network to satisfy a certain traffic demand. As observed in Fig. 1, the inputs of the model are the traffic demand (either actual or predicted), the average spectral efficiency and the candidate network configuration.

In the model, the transmit power $\hat{P}_i^{(t)}$ for each SC of the candidate solution is determined by:

$$\hat{P}_i^{(t)} = P_N \cdot G_{PL,i}^{(t)}(u_{edge}^{(t)}) \cdot \left|\hat{F}_i^{(t)}\right| \cdot SINR_{edge} \quad (15)$$

where $P_N$ is the noise power measured in one channel, $G_{PL,i}^{(t)}(u_{edge}^{(t)})$ is the path gain (loss) from SC $i$ to pixel $u_{edge}^{(t)}$ located at the cell-edge and $SINR_{edge}$ represents the target value at such a distance. The range of $P_i^{(t)}$ is limited by the maximum allowed transmit power, $P_{max}$. The cell-edge is a function of the inter-site distance (ISD), which is given by the distance to the closest adjacent SC.

The received power $\hat{P}_{RX,i}^{(t)}(u)$ at pixel $u$ when served by SC $i$ using a single channel is given by:

$$\hat{P}_{RX,i}^{(t)}(u) = \hat{P}_i^{(t)} \cdot G_i(u). \quad (16)$$

where $G_i(u)$ is the overall gain between SC $i$ and pixel $u$, expressed as the sum of individual gains (losses) including the antenna gain and the path loss.

The users are served by the SC from which they receive the strongest received power. In this way, the function $\hat{\Gamma}^{(t)}(u)$ returns the SC that serves the users in pixel $u$ (i.e. it defines the service area of every SC). Formally, it is defined as:

$$\hat{\Gamma}^{(t)}(u) = \arg\max_i \hat{P}_{RX,i}^{(t)}(u). \quad (17)$$

Such a function facilitates the conversion between the SC and pixel domains in the model.

The $SINR_{i,k}^{(t)}(u)$ at pixel $u$ when served by SC $i$ using channel $k$ is expressed as:

$$SINR_{i,k}^{(t)}(u) = \frac{\hat{P}_{RX,i}^{(t)}(u)}{\left(\sum_{j \in U_S \setminus \{i\}} \bar{\alpha}_j^{(t)} \cdot \pi_j^{(t)}(k) \cdot \hat{P}_{RX,j}^{(t)}(u)\right) + P_N}, \quad (18)$$

where $\pi_j^{(t)}(k)$ indicates whether channel $k$ is allocated to SC $j$ (with value 1) or not (0) and $\bar{\alpha}_j^{(t)}$ is the average load. Computing the SINR requires to solve a system of non-linear equations due to the load-coupling, i.e. the load of a cell is a function of the load levels of other cells [38]. To simplify this procedure, the average load $\bar{\alpha}_j^{(t)}$ in SC $j$ is approximated by:

$$\bar{\alpha}_j^{(t)} \approx \min\left(\frac{D_j^{(t)}}{C_j^{(t-1)}}, 1\right), \quad (19)$$

where $C_j^{(t-1)}$ is the capacity of SC $j$ at the previous time step. The average $SINR_i(u)$ at pixel $u$ when served by SC $i$ is given by:

$$SINR_i^{(t)}(u) = \frac{1}{|\hat{F}_i^{(t)}|} \sum_{k \in \hat{F}_i^{(t)}} SINR_{i,k}^{(t)}(u), \quad (20)$$

Then, the spectral efficiency $SE_i^{(t)}(u)$ at pixel $u$ is obtained by:

$$SE_i^{(t)}(u) = f_{RAT}(SINR_i^{(t)}(u)) \quad (21)$$

where $f_{RAT}(\cdot)$ is a function that depends on the radio access technology (e.g. LTE). From the spectral efficiency and the traffic demand at the pixel-level, the required bandwidth $B_{i,u}'^{(t)}$ is calculated based on (3). Lastly, this information can be aggregated on a cell basis using the function $\hat{\Gamma}^{(t)}(u)$ as follows:

$$B_i'^{(t)} = \sum_{u | i = \hat{\Gamma}^{(t)}(u)} B_{i,u}'^{(t)}. \quad (22)$$

## V. PERFORMANCE EVALUATION

### A. Simulation scenario

An urban SC scenario with dimensions 0.4 km × 0.4 km and a grid resolution of 5 m has been considered. To represent the areas where deploying SCs is possible, e.g. no backhaul and site acquisition constraints, 2% of the points (or pixels) in the scenario have been randomly selected as candidate site locations. The actual network layout and the traffic demand at the busy hour in the situation before the consideration of the new tenant are represented in Fig. 3, where the triangles represent the location of the three deployed SCs and the values in brackets are the number of allocated channels. Color scale indicates the traffic demand density, which is non-uniformly distributed over the considered area. The traffic demands supported by SCs 1-3 are 8.8, 5.6 and 5.0 Mbps, respectively.

The *network performance model* has been implemented according to Section IV.B. Table I summarizes the main parameters of this model. The transmit power $P_i^{(t)}$ is configured for each SC to have $SINR_{edge} = 9$ dB at $\frac{\sqrt{3}}{2}$ of the

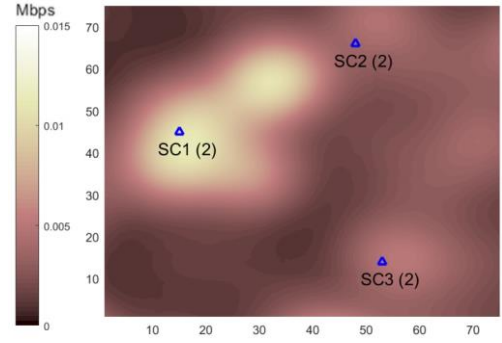

*Fig. 3. Traffic demand and network deployment in the initial situation (before new tenant's arrival).*

ISD [36]. The spectral efficiency function $SE(SINR)$ used to compute the average spectral efficiency $\overline{SE}_i^{(t)}$ at SC $i$ depending on the SINR at each pixel is obtained from Section A.1 in [39] with $SE_{max} = 4.4$ b/s/Hz.

| Parameter | Configuration |
|---|---|
| Deployment scenario | Urban, small cells, 0.4km x 0.4km |
| Operating frequency | 5 GHz |
| Channel bandwidth | 20 MHz |
| Cell bandwidth | 4 channels |
| Propagation (path loss) | ITU InH model [40] |
| SC antenna directivity | omni-directional |
| SC antenna height | 6 m |
| UE antenna height | 1.5 m |
| SC antenna gain | 2 dBi |
| UE thermal noise | -174 dBm/Hz |
| UE noise figure | 9 dB |
| UE minimum SINR | -10 dB |
| SC TX power range | [10-24] dBm |

TABLE I. SIMULATION PARAMETERS

From a network planning perspective, the parameters used in the *capacity conformance monitoring* module to trigger the planning actions are configured as: $\alpha = 0.95$, $\beta = 0.7$ and $\gamma = 0.05$. Parameter $\alpha$ is configured assuming moderate traffic variations over the expected values. However, depending on the provider's deployment policies, this parameter can be configured with a lower value in order to provide higher levels of spare capacity in the SCs and thus leaving some room for coping with unexpected traffic variations. Regarding parameters $\beta$ and $\gamma$, a reasonable configuration of such parameters has been considered in this work to react to traffic variations while, at the same time, limiting the number of "re-planning" actions and targeting an efficient resource utilization. In addition, the maximum number of allocated channels per SC is set to $K_{max} = 3$, while the maximum number of SCs that can be deployed in the considered area is set to $N_{max}^{SC} = 10$.

### B. Analysis of the network planning solutions

Let assume that the SLA of the new tenant is translated to a specification at the busy hour of $A_m^{(t_B)} = 100$ Mbps. At this initial stage, the new tenant's spatial traffic demand distribution is assumed to be unknown, so the planning is carried out using the detailed planning specifications from the methods explained in Section IV.A. Thus, the total traffic demand is calculated as the actual traffic demand from existing tenants plus the estimated new tenant's demand. After generating the detailed planning specifications, it is observed that in the *capacity conformance monitoring* module condition (14) is satisfied for the three deployed SCs, so that the *capacity dimensioning and planning* module is launched.

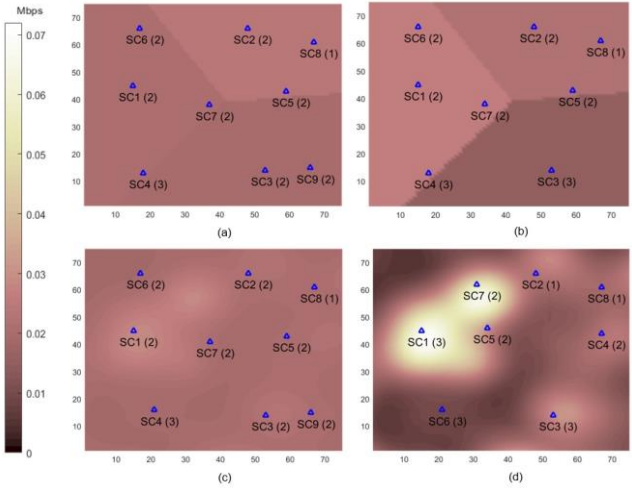

Fig. 4. Network deployment and estimated traffic demand using the detailed planning specifications: (a) Based on uniform distribution at the SC-level, (b) Based on correlated distribution at the SC-level, (c) Based on uniform distribution at the pixel-level, and (d) Based on correlated distribution at the pixel-level.

Fig. 4(a-d) show the results of the planning process for different sets of detailed planning specifications corresponding to the methods of Section IV.A.

For the uniform distribution at the SC-level method, Fig. 4(a) shows that, given that SC 2 has the smallest service area, the traffic demand per pixel in this SC is estimated to be slightly higher than in SC 1 and 3. Then, the *capacity dimensioning and planning* module adds 6 new SCs, three of which are located in the right upper corner of the scenario, where the traffic density is higher.

The correlated distribution at the SC-level method is represented in Fig. 4(b). This method estimates that SC 3, which initially carried less traffic (i.e. 5.0 Mbps), is the cell that receives proportionally less traffic from the new tenant. Therefore, compared to the method of Fig. 4(a), additional SCs such as SC 9 are not required in the service area of SC 3. Instead, the number of channels allocated to SC 3 is increased by one.

For the method based on uniform distribution at the pixel-level, illustrated in Fig. 4(c), the results are quite similar to the first method since both methods approximate the uniform distribution. The only difference is that SC 4 is placed a bit more to the right.

The last method [see Fig. 4(d)], based on correlated distribution at the pixel-level, produces the largest variations in the traffic demand per pixel. Consequently, SCs 5 and 7 are placed in, or close to, the area with high traffic density, so that part of the traffic is offloaded from SC 1, having this cell three channels allocated after the planning. Besides, unlike other methods, additional SCs are not required in the service area of SC 3 because a lower traffic density is assumed in this region.

C. *Analysis of the network operation with the new tenant*

This section evaluates the solutions of the planning algorithm when the new tenant's service is operative and the actual traffic demand at the busy hour of the new tenant is spatially distributed as illustrated in Fig. 5, where two cases are distinguished. In the former [Fig. 5(a)], the new tenant's spatial traffic demand exhibits quite high correlation with already existing tenants, whose spatial traffic distribution is represented in Fig. 3. Specifically, using Pearson's coefficient, both traffic distributions are 90% correlated. In the latter case [Fig. 5(b)], the distributions are only 15% correlated.

Let assume now that the network has been deployed as dictated by the planning [i.e. with the real network layouts as illustrated in Fig. 4(a-d)] and let consider the real traffic demand of the new tenant shown in Fig. 5. In that case, Table II shows the required bandwidth $\tilde{B}_i^{(t)}$ and the cell bandwidth $B$ in each SC considering the actual traffic demand for the two levels of correlation with the different planning methods. The last row in the table shows values aggregated over all the SCs. The notation in the table is X/Y where X represents the required bandwidth and Y the cell bandwidth. As a reference for comparison with the methods discussed in Fig. 4, the table also includes the result of the network planning taking as input the real traffic of the new tenant (shown in Fig. 5). The deployments for this case are shown in Fig. 6.

In general, the method that fits better the traffic demand (in this case, the reference approach) will minimize the required resources without generating a loss of traffic. However, according to Table II, this does not necessarily mean a lower value of total required bandwidth. For example, the methods based on uniform distribution provide the lowest values; however, this happens because these methods deploy a greater number of SCs in the scenario, as reflected in Fig. 4(a) and (c).

Given the minimum number of deployed SCs (i.e. 8), the reference case obtains the lowest value of total required bandwidth. With respect to the methods based on correlated distribution, the method with SC-level resolution results in a lower total required bandwidth since its network layout is more similar to the reference case, as previously stated. Regarding the two methods based on uniform distribution, the results in terms of required bandwidth are very close to each other because of the similarity of their network layouts.

Another aspect from Table II (see numbers highlighted in bold) is that the required bandwidth in some SCs exceeds (or nearly exceeds) the cell bandwidth, meaning that some traffic might be lost. These cases are more evident in the case of 15% correlated traffic due to the poorer match between the network layout and the spatial distribution of traffic demand. In the case of 90% correlated traffic, only the methods based on correlated distribution provide insufficient bandwidth or they are close to it. However, this lack of bandwidth (about 1 MHz) is marginal compared to the channel bandwidth. In addition, the deployment has been carried out with only 8 SCs, so that the cost of the solution is cheaper than other methods.

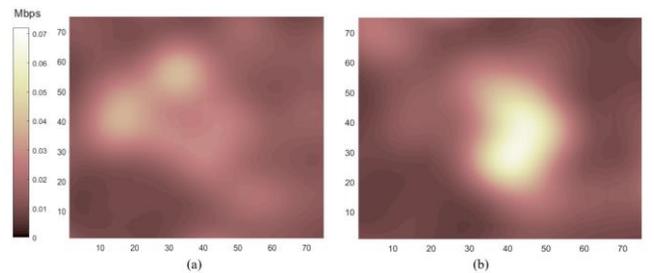

Fig. 5. Traffic demand of the new tenant: (a) 90% correlated with network's traffic demand; (b) 15% correlated with network's traffic demand.

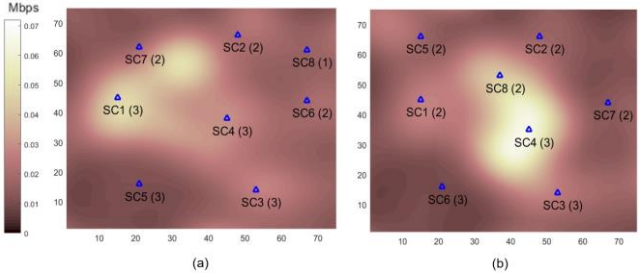

*Fig. 6. Network deployment with real tenant's traffic demand for: (a) 90% correlated traffic; (b) 15% correlated traffic.*

| Cor. [%] | SC | REFERENCE (ACTUAL TRAFFIC KNOWN) | UNIFORM SC-LEVEL | CORR. SC-LEVEL | UNIFORM PX-LEVEL | CORR. PX-LEVEL |
|---|---|---|---|---|---|---|
| 90 | 1 | 30/60 | 32/40 | 30/40 | 29/40 | 38/60 |
|    | 2 | 18/40 | 23/40 | 28/40 | 19/40 | 14/20 |
|    | 3 | 43/60 | 25/40 | 48/40 | 24/40 | 53/60 |
|    | 4 | 28/60 | 23/40 | 23/40 | 32/60 | 24/40 |
|    | 5 | 35/60 | 17/40 | 29/40 | 19/40 | **41/40** |
|    | 6 | 16/40 | 24/40 | 22/40 | 22/40 | 37/60 |
|    | 7 | 36/40 | 36/40 | **39/40** | 36/40 | 28/40 |
|    | 8 | 9/20  | 10/20 | 10/20 | 10/20 | 8/20 |
|    | 9 | --    | 10/40 | --    | 10/40 | -- |
|    | tot | 215/380 | 200/360 | 229/340 | 201/360 | 243/340 |
| 15 | 1 | 24/40 | 21/40 | 19/40 | 19/40 | 38/60 |
|    | 2 | 22/40 | 20/40 | 24/40 | 16/40 | 15/20 |
|    | 3 | 42/60 | 31/40 | 53/60 | 33/40 | **66/60** |
|    | 4 | 28/60 | 21/60 | 20/60 | 30/60 | 30/40 |
|    | 5 | 14/40 | 19/40 | 35/40 | 22/40 | **69/40** |
|    | 6 | 30/60 | 23/40 | 20/40 | 22/40 | 36/60 |
|    | 7 | 27/40 | **48/40** | **55/40** | **49/40** | 38/40 |
|    | 8 | 19/40 | 7/20 | 7/20 | 7/20 | 6/20 |
|    | 9 | --    | 9/40 | --   | 9/40 | -- |
|    | tot | 206/380 | 199/360 | 233/340 | 207/360 | 298/340 |

TABLE II. ACTUAL REQ. BW [MHZ] AND CELL BW [MHZ] FOR 90 AND 15% CORRELATED TRAFFIC

### D. Re-planning the new tenant during tenant's operation

Once the new tenant's service is operative, condition (14) is evaluated again to determine whether there exists a lack of capacity or not. If so, the *capacity dimensioning and planning* module is relaunched to provide a new network configuration. In our experiment, this happens for the SCs whose statistics in Table II are represented in bold.

For the methods based on uniform distribution, it is noted that, in case of 90% correlated traffic, there is no lack of capacity. However, in case of 15% correlated traffic, SC 7 satisfies condition (14) and therefore a new planning stage is launched. As a result of the re-planning process, it is obtained that the network layout is not modified, but SCs 4 and 7 increase the number of channels by one, while SC 9 decreases it by one.

With respect to the methods based on correlated distribution, Fig. 7 shows the network layout after the re-planning process for 90% and 15% correlated traffic. As observed, both the network layout and bandwidth assignment have changed. In case of 90% correlated traffic [Fig. 7(a) and (b)], a new SC (i.e. SC9) is deployed in the left upper side of the scenario to relieve traffic from congested SCs. This new SC also produces changes in the bandwidth assignment, which can be observed by comparing the numbers in parentheses in Fig. 4(b) and (d) with those of Fig. 7(a) and (b), respectively. In total, there are three changes (i.e. adding or removing a channel) in each case.

In case of 15% correlated traffic [Fig. 7(c) and (d)], a new SC (i.e. SC9) is deployed in the center of the scenario, where the traffic density is higher. With respect to the channel assignment, the changes can be observed by comparing Fig. 4(b) and (d) with Fig. 7(c) and (d), respectively.

Table III shows a comparative analysis between the network layouts before and after the re-planning process for each analyzed method. For a high level of correlated traffic (90%), the best methods (without considering the reference) are the two based on correlated distribution, since they utilize the lowest number of channels (i.e. 15 and 17), provided that the number of deployed SCs is 9 for all methods. Note that these two solutions are achieved through a two-step process that comprises planning and re-planning (where an additional SC is deployed). If the method employs pixel-level resolution, there is also a bandwidth shortage (about 1 MHz), which might lead to a small loss of traffic. For this reason, the method based on correlated distribution at the SC-level is a better solution when the new tenant's traffic is not fully correlated with already existing tenants.

For a low level of correlated traffic (15%), it is observed that all the methods result in bandwidth shortage before the re-planning stage, since the network layouts do not fit properly the traffic demand. Such an effect is more pronounced for the methods based on correlated distribution, especially when pixel-level resolution is used. In this latter case, because of the bad traffic estimation, the number of channels assigned after re-planning represents the worst case (i.e. 20). In case of the SC-level resolution, there is a bandwidth shortage of 15 MHz. However, this method eliminates the lack of bandwidth by adding a new SC in an optimal location during the re-planning phase, while the other methods that previously deployed more SCs are unable to improve the solution.

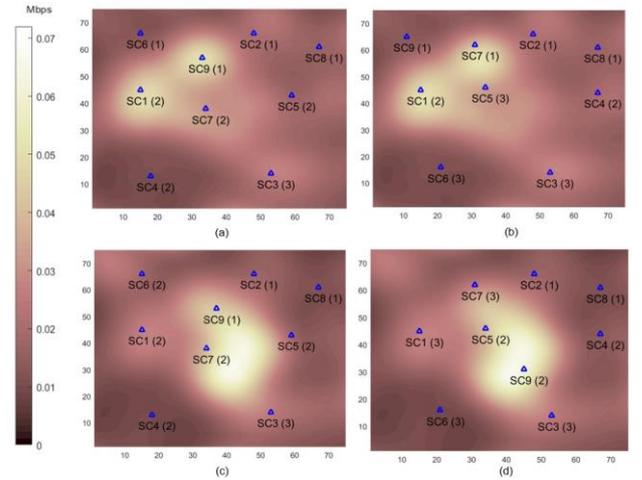

*Fig. 7. Network deployment with real tenant's traffic demand for: (a) 90% correlated traffic, method based on correlated distribution at the SC-level; (b) 90% correlated traffic, method based on correlated distribution at the pixel-level; (c) 15% correlated traffic, method based on correlated distribution at the SC-level; (d) 15% correlated traffic, method based on correlated distribution at the pixel-level.*

It is clear that the pixel-level methods do not leverage the higher spatial resolution when the traffic correlation is low, as they have to employ the greatest number of channels. Thus, the best methods (excluding the reference) in terms of minimum number of channels are the two methods with SC-level resolution. An important difference between them is the amount of traffic that could be lost before the re-planning stage. Thus, although one method requires less amount of resources, it might lead to higher traffic losses.

| Cor. [%] | Method | #SCs | | #channels | | Req. BW [MHz] | | BW shortage [MHz] | |
|---|---|---|---|---|---|---|---|---|---|
| | | Bef | Aft | Bef | Aft | Bef | Aft | Bef | Aft |
| 90 | REFERENCE | 8 | 8 | 19 | 19 | 215 | 215 | 0 | 0 |
| | UNIFORM SC-LEVEL | 9 | 9 | 18 | 18 | 200 | 200 | 0 | 0 |
| | CORR. SC-LEVEL | 8 | 9 | 17 | 15 | 229 | 210 | 0 | 0 |
| | UNIFORM PX-LEVEL | 9 | 9 | 18 | 18 | 201 | 201 | 0 | 0 |
| | CORR. PX-LEVEL | 8 | 9 | 17 | 17 | 243 | 215 | 1 | 0 |
| 15 | REFERENCE | 8 | 8 | 19 | 19 | 206 | 206 | 0 | 0 |
| | UNIFORM SC-LEVEL | 9 | 9 | 18 | 17 | 199 | 205 | 8 | 0 |
| | CORR. SC-LEVEL | 8 | 9 | 17 | 16 | 233 | 215 | 15 | 0 |
| | UNIFORM PX-LEVEL | 9 | 9 | 18 | 18 | 207 | 198 | 9 | 0 |
| | CORR. PX-LEVEL | 8 | 9 | 17 | 20 | 298 | 228 | 35 | 0 |

TABLE III. NETWORK DEPLOYMENT BEFORE AND AFTER RE-PLANNING

### E. Comparison with the state-of-the-art

As explained in Section II, various approaches to solve the RNP problem have been proposed in the literature. In this paper, the selected approach, summarized in Algorithm 1, is based on an iterative approach where a certain planning action (such as deploying a new SC or adding a channel) is executed at each step. Previous works based on iterative approach [20-21] reduce the set of planning actions at each step to determining the location of the SC ([20]) and, optionally, in heterogeneous networks, selecting the optimal bandwidth allocation with respect to the macrocell layer ([21]). However, the problem of bandwidth allocation in the SC layer (i.e. with respect other SCs) has not been addressed in those works.

Based on the above considerations, a state-of-the-art (SOTA) method has been developed to compare the performance with the proposed Algorithm 1. Specifically, the SOTA method implements the iterative approach in [20-21]. Unlike Algorithm 1, this method only comprises the planning action of deploying a new SC. Since the actions of adding or removing a channel are not available during the planning process, the number of allocated channels per SC must be constant. Lastly, channel selection is performed according to Algorithm 2.

The SOTA method has been evaluated under two distinct contexts. One takes as input the real traffic of the new tenant in the same way that Algorithm 1 was evaluated as a reference in Section V.C (see Table II). The other combines Algorithm 1 with the best planning method of Section IV.A used to derive the detailed planning specifications. According to the evaluations in Section V.A-D, the best solution corresponds to the method based on correlated distribution at the SC-level, since it employed the least amount of network resources. The study has been performed with two different values of the number of channels per SC, i.e. 2 and 3 channels. This constraint is only applied to the newly deployed SCs in the scenario, because the SOTA method does not consider the possibility of changing the number of allocated channels in existing SCs. In addition, evaluations are carried out for the two levels of traffic correlation (90 and 15%) used in previous sections.

Table IV shows the required bandwidth and the cell bandwidth in each SC for the SOTA method under the above-explained conditions (the notation is the same as in Table II). The values in the table correspond to the situation when the updated network is operative and carries actual traffic from the new tenant. Compared to Algorithm 1 (see Table II), it is observed that, in general, the number of SCs is larger with the SOTA method. In addition, since the same (constant) bandwidth is allocated for all the newly deployed SCs, the total number of allocated channels is also larger. This highlights that the possibility of changing the number of channels per SC (e.g. by considering a larger set of possible planning actions to choose from as in Algorithm 1) is much more effective than limiting the cell bandwidth to a constant value and just considering the addition of new SCs as in the SOTA method. A closer look at Table IV reveals that, when three channels per SC are allocated, the SOTA method based on correlated distribution at the SC-level employs, as expected, a larger amount of resources (10 SCs) than if the actual traffic is known (8-9 SCs). However, when the cell bandwidth is limited to two channels, the amount of resources in both cases is the same (10 SCs). In turn, the method proposed in this paper is able to support the traffic with only 8 SCs (see Table II), thus outperforming the SOTA method with both 2 and 3 channels.

### VI. CONCLUSION

In this paper, the cell planning problem for small cell multi-tenant networks has been studied. From the perspective of infrastructure providers, the automation of procedures is a key consideration due to the complexity of managing diverse tenants' capacity requirements. In the proposed scheme, these requirements are translated into a set of detailed planning specifications over the spatial/temporal domains. Then, the planning process is modeled following a SON approach, where a condition to detect capacity issues is periodically checked in order to trigger particular planning actions, such as adding/removing a channel or deploying/relocating a SC.

The proposed framework has been evaluated in a scenario in which a new tenant is added in the network. To derive the set of planning specifications of the new tenant, different methods are considered depending on the expected correlation with the actual traffic demand in the network and the spatial resolution of the traffic measurements. The evaluation has been carried out for two different traffic correlation levels. Results show that the detailed planning specifications based on correlated distribution with a spatial resolution at the SC-level employ the least amount of network resources. This is because the differences between the estimated and actual traffic demand make the use of higher spatial resolutions less effective. The specifications based on uniform distribution require a larger amount of resources to meet the traffic demand even if the new tenant's traffic and network's traffic are poorly correlated. In addition, the proposed capacity and dimensioning scheme (Algorithm 1) has been compared with existing planning solutions, which do not consider the spectrum planning in the SC-layer. Results show that the existing solutions require a larger number of deployed SCs for serving the same traffic than the proposed approach.

| Cor. [%] | SC | SOTA-2CH (ACTUAL TRAFFIC KNOWN) | SOTA-3CH (ACTUAL TRAFFIC KNOWN) | SOTA-2CH + CORR. SC-LEVEL | SOTA-3CH + CORR. SC-LEVEL |
|---|---|---|---|---|---|
| 90 | 1 | 27/40 | 30/40 | 20/40 | 23/40 |
|  | 2 | 13/40 | 22/40 | 26/40 | 24/40 |
|  | 3 | 20/40 | 22/40 | 12/40 | 14/40 |
|  | 4 | 16/40 | 16/60 | 19/40 | 16/60 |
|  | 5 | **45/40** | 34/60 | 23/40 | 17/60 |
|  | 6 | 28/40 | 22/60 | 25/40 | 18/60 |
|  | 7 | 9/40 | 22/60 | 33/40 | 27/60 |
|  | 8 | 8/40 | 14/60 | 8/40 | 6/60 |
|  | 9 | 8/40 | 16/60 | 8/40 | 9/60 |
|  | 10 | 19/40 | -- | 10/40 | 8/60 |
|  | tot | 193/400 | 198/480 | 184/400 | 162/540 |
| 15 | 1 | 22/40 | 31/40 | 13/40 | 15/40 |
|  | 2 | 8/40 | 38/40 | 23/40 | 20/40 |
|  | 3 | 23/40 | 25/40 | 17/40 | 19/40 |
|  | 4 | 31/40 | 40/60 | 15/40 | 13/60 |
|  | 5 | **45/40** | 22/60 | 26/40 | 20/60 |
|  | 6 | 17/40 | 35/60 | 24/40 | 17/60 |
|  | 7 | 7/40 | 16/60 | **44/40** | 35/60 |
|  | 8 | 8/40 | 8/60 | 6/40 | 4/60 |
|  | 9 | 14/40 | -- | 10/40 | 12/60 |
|  | 10 | 12/40 | -- | 10/40 | 8/60 |
|  | tot | 187/400 | 215/420 | 188/400 | 163/540 |

TABLE IV. ACTUAL REQ. BW [MHZ] AND CELL BW [MHZ] FOR SOTA METHOD

As future work, it is planned to further analyze the proposed planning methodology for ongoing tenants whose traffic demand varies significantly over time. In particular, when the traffic demand of a certain tenant decreases, planning actions such as channel releases and SC relocations become effective solutions to minimize capacity over-provisioning in the network. In addition, more sophisticated combinatory optimization will be investigated for solving the problem.

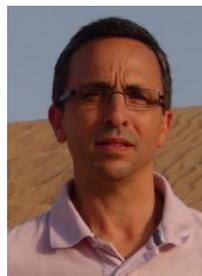

**Oriol Sallent** is a Professor at the Universitat Politècnica de Catalunya (UPC) in Barcelona. He has participated in a wide range of european and national projects, with diverse responsibilities as Principal Investigator, Coordinator and Workpackage Leader. He regularly serves as a consultant for a number of private companies. He has been involved in the organization of many different scientific activities, such as Conferences, Workshops, Special Issues in renowed international journals, etc. He has contributed to standardisation bodies such as 3GPP, IEEE and ETSI.

He is co-author of 13 books and has published 200+ papers, mostly in high-impact IEEE journals and renowed international conferences. His research interests include 5G RAN (Radio Access Network) planning and management, artificial intelligence-based radio resource management, virtualisation of wireless networks, cognitive management in cognitive radio networks and dynamic spectrum access and management among others.

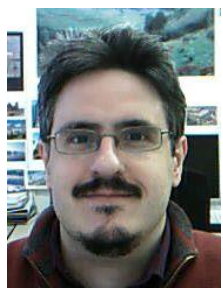

**Jordi Pérez-Romero** (S'98–M'04) is associate professor at the Dept. of Signal Theory and Communications of the Universitat Politècnica de Catalunya (UPC) in Barcelona, Spain. He received the Telecommunications Engineering degree and the PhD from the same university in 1997 and 2001, respectively. His research interests are in the field of mobile communication systems with a main focus of 5G, covering radio resource and QoS management, self-organizing networks, network slicing, multi-tenancy and application of data analytics and artificial intelligence tools in the management of 5G networks. He has been involved in different European Projects as well as in projects for private companies. He has published more than 200 papers in international journals and conferences and has co-authored two books on mobile communications. He is associate editor of IEEE Vehicular Technology Magazine and Eurasip Journal on Wireless Communications Networks.

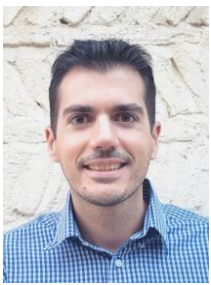

**Pablo Muñoz** received his M.Sc. and Ph.D. degrees in telecommunication engineering from the University of Málaga in 2008 and 2013, respectively. From 2009 to 2013, he was a Ph.D. fellow in self-optimization of mobile radio access networks. Upon completing his Ph.D, he worked as a research assistant within an R&D contract with Optimi-Ericsson. Since 2016 he is the recipient of a Spanish competitive grant (Juan de la Cierva) at the Polytechnic University of Catalonia.